\documentclass{article}
\usepackage{spconf,amsmath,graphicx,booktabs}
\usepackage{fancyhdr}

\ninept

\title{Autoencoder with Group-based Decoder and Multi-task Optimization for Anomalous Sound Detection}
\name{Yifan Zhou$^1$, Dongxing Xu$^2$, Haoran Wei$^3$, Yanhua Long$^{1,2*}$\thanks{Yanhua Long is the corresponding author, she is also with the Shanghai Engineering Research Center of Intelligent Education and Bigdata, Shanghai Normal University, Shanghai, China. The work is supported by the National Natural Science Foundation of China (Grant No.62071302).}}

\address{
  $^1$Key Innovation Group of Digital Humanities Resource and Research, \\
  Shanghai Normal University, Shanghai, China\\
  $^2$Unisound AI Technology Co., Ltd., Beijing, China\\
  $^3$Department of ECE, University of Texas at Dallas, Richardson, TX 75080, USA}
\begin{document}




\maketitle

\thispagestyle{fancy}
\fancyhead{}
\lhead{}
\lfoot{\copyright2024 IEEE. Personal use of this material is permitted. Permission from IEEE must be obtained for all other uses, in any current or future media, including reprinting/republishing this material for advertising or promotional purposes, creating new collective works, for resale or redistribution to servers or lists, or reuse of any copyrighted component of this work in other works.}
\cfoot{}
\rfoot{}

\begin{abstract}
   In industry, machine anomalous sound detection (ASD) is in great demand. However, collecting enough abnormal samples is difficult due to the high cost, which boosts the rapid development of unsupervised ASD algorithms. Autoencoder (AE) based methods have been widely used for unsupervised ASD, but suffer from problems including ‘shortcut’, poor anti-noise ability and sub-optimal quality of features. To address these challenges, we propose a new AE-based framework termed AEGM. Specifically, we first insert an auxiliary classifier into AE to enhance ASD in a multi-task learning manner. Then, we design a group-based decoder structure, accompanied by an adaptive loss function, to endow the model with domain-specific knowledge. Results on the DCASE 2021 Task 2 development set show that our methods achieve a relative improvement of 13.11\% and 15.20\% respectively in average AUC over the official AE and MobileNetV2 across test sets of seven machines.
\end{abstract}

\begin{keywords}  
   Anomalous sound detection, autoencoder, multi-task optimization, group-based decoder
\end{keywords}


\vspace{1em}
\section{Introduction}
\begin{figure*}[ht]
   \centering
   \includegraphics[width=0.85\linewidth]{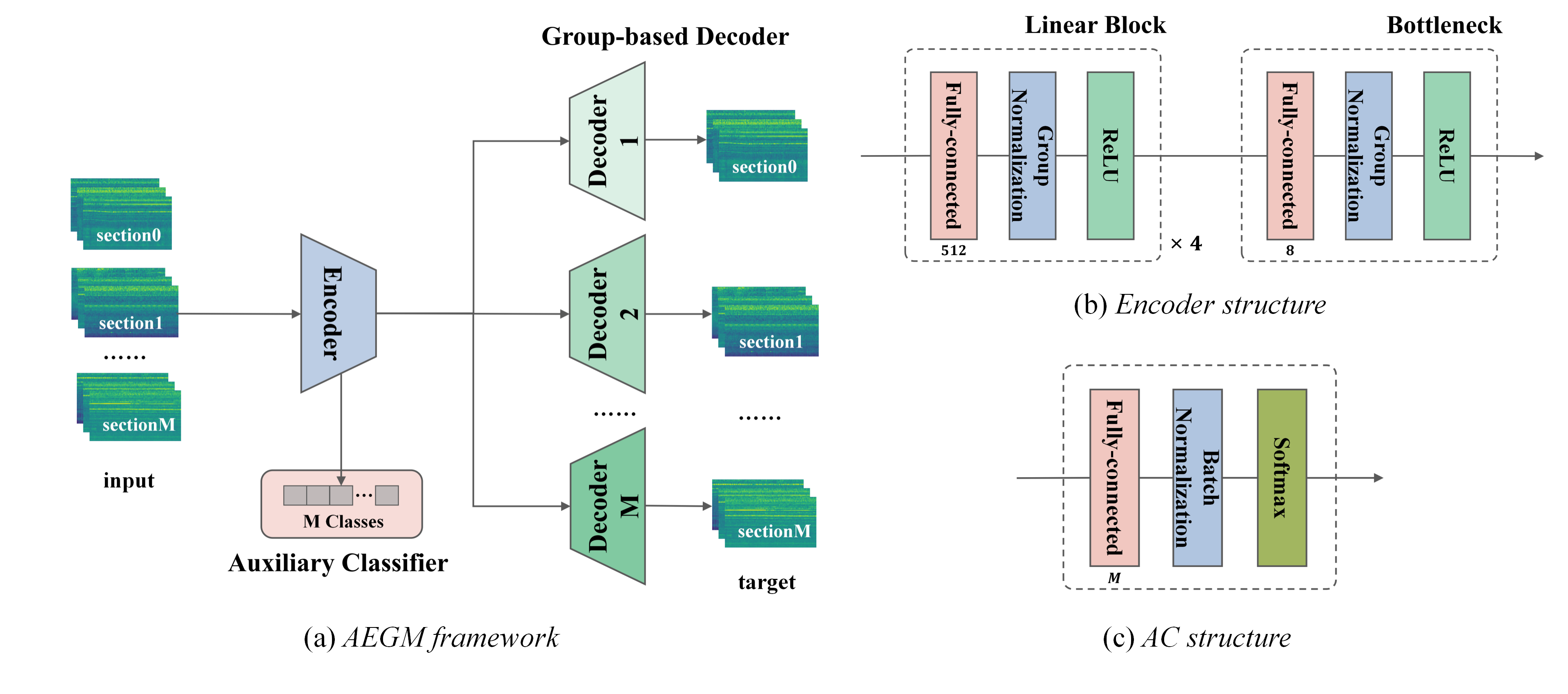}
   \caption{The framework of proposed AEGM. Different sections represent different machine IDs (groups, categories) of each ASD task. The encoder-decoder structure is symmetrical. The auxiliary classifier performs inter-machine classification of normal samples. }
   \label{fig:overview}
 \end{figure*}

In recent years, with the launch of Detection and Classification of
Acoustic Scenes and Events (DCASE) challenges \cite{Kawaguchi_arXiv2021_01}, the
anomalous sound detection (ASD) \cite{koizumi2018unsupervised}
technologies for industrial machine condition monitoring have
attracted increasing research attention. However, anomalous sounds rarely occur and are highly diverse in real-world scenarios. Collecting a large number of anomalous samples is extremely costly and nearly impossible \cite{huang2018internet}. In addition,
because of the complicated variations of machine working conditions,
there is a large domain shift between the training and testing phases \cite{Kawaguchi_arXiv2021_01}.
Therefore, building domain-robust unsupervised ASD systems with only
normal sound clips becomes very necessary.

Although the published works on ASD are still limited in recent years, there are
many previous works related to other anomaly detection (AD) methods in the literature.
The mainstream AD algorithms can be classified into the following
three types: 1) reconstruction-based methods \cite{marchi2015novel,baldi2012autoencoders,kingma2013auto,sabokrou2018adversarially},
which assume that a model trained on normal samples will have a high
reconstruction error when applied to abnormal samples;
2) feature-learning-based methods\cite{hendrycks2018deep,ruff2020rethinking,chen2020simple,hojjati2022self},
which extract distinctive features through proxy tasks such as normal sample
classification or contrastive learning to perform anomaly detection;
3) distribution-based methods \cite{purohit2020deep,scott2004outlier,breunig2000lof,liu2008isolation},
which model the distribution of normal samples and perform anomaly detection.
In summary, while various AD methods have been developed, the reconstruction-based autoencoder (AE) has become widely used due to their simplicity and effectiveness.
For example, \cite{an2015variational} proposed to use the reconstruction
probability from variational AE to improve the AD performance;
Authors in \cite{suefusa2020anomalous} proposed the use of interpolation deep neural network,
which employs the reconstruction error between predicted and actual frames in the spectrogram
to detect anomalies, resulting in a significant improvement
in detecting non-stationary machine sounds as compared to standard AE.
In \cite{purohit2022hierarchical}, authors used the hierarchical conditional
variational AE to incorporate a hierarchy of prior knowledge,
such as machine category or IDs, to refine the representation of the latent space.
While in \cite{georgescu2021background}, authors proposed a method that
utilizes adversarial training to generate pseudo anomaly samples,
which alleviates the anomalous data sparsity at some extent and makes
the AEs background-agnostic.

Although extensive studies have explored reconstruction-based AEs for AD, challenges remain limiting their unsupervised ASD performance. First, AE tends to reconstruct abnormal samples,
primarily because it learns ‘shortcut’ that the model copies and pastes information,
which can lead to failure in anomaly detection \cite{you2022unified}. Second, AE trained only
on normal samples not only results in poor quality of deep embeddings, but also
leads to poor model ‘noise’ robustness \cite{pang2021deep}.
In this study, to
deal with these challenges, we propose a new simple AE-based framework, called AEGM,
to improve the performance of ASD tasks. Our contributions are: 1)
We construct an auxiliary classifier in AE with multi-task optimization
to alleviate the ‘shortcut’ problem, by leveraging the discriminative
information of encoder bottleneck embeddings between different
types of normal samples;  2) A group-based decoder structure with an adaptive loss function is
designed to enable the AE with domain specialization ability.
Our experiments use the DCASE 2021 Task 2 development set. Results show the proposed methods significantly outperform official baselines and other AE-based methods across seven machine ASD test sets.

\vspace{1em}
\section{Proposed Method}


The proposed AEGM framework consists of an encoder, a simple auxiliary classifier and a group-based decoder. This framework is illustrated in Fig.\ref{fig:overview}(a).
The design inspiration of AEGM comes from ‘stirring up muddy water causes sand to rise to the surface’. It enhances anomaly exposure by ‘stirring’ the training samples through joint training of the group-based AE and auxiliary classifier based on reconstruction and classification criterions, respectively. Details of AEGM are presented below.

\vspace{1em}
\subsection{Auxiliary Classifier}

It is well known that ‘shortcut’ is a typical issue of AE-based methods.
When ‘shortcut’ occurs, the model copies the sample in the encoder
and pastes it in the decoder. At this point, during inference,
anomalous samples can also be reconstructed, resulting in a failure of anomaly detection.
Taking the following fully connected single layer with weights $\boldsymbol{w}$ and biases $\boldsymbol{b}$ as an example,
\begin{equation}
  \boldsymbol{y}=\boldsymbol{w}\boldsymbol{x}+\boldsymbol{b}
  \label{eq1}
\end{equation}
when $\boldsymbol{w}$ is set to $\boldsymbol{I}$ (identity matrix) and $\boldsymbol{b}$ is set to $\boldsymbol{0}$ (zero matrix), the model will exhibit a ‘shortcut’ and lead to a failure in detection\cite{you2022unified}.


To alleviate the ‘shortcut’ problem, we propose an auxiliary classifier (AC) in the AE. Motivated by the outlier exposure idea in \cite{hendrycks2018deep}, the AC performs a classification task in the latent space to ‘stir’ the training samples and promote outlier exposure.
For computational efficiency, this classifier contains only one linear layer followed by softmax. Its detailed structure is shown in Fig.\ref{fig:overview}(c).
The AC classifies different machine sections (IDs, groups, categories) based on the representations learned by the AEGM encoder. By taking the encoder's bottleneck embeddings as input, the proposed AC is trained with the cross entropy loss function as
\begin{equation}
  \mathcal{L}_{\text{aux}}=-\frac{1}{NM}\sum_{i=1}^{N} \sum_{j=1}^{M} y_{ij} \log \left(p_{\theta_{E}}\left(x_{i}\right)\right)
  \label{eq2}
\end{equation}
where $N$ and $M$ mean the total input samples of each section and total sections of each machine, $x_{i}$,
$y_{ij}$ represent the $i$-th input sample and its target of the $j$-th section
of the machine. $p_{\theta_{E}}(\cdot)$ is the softmax output of AC with parameters $\theta_{E}$.

Like the confidence score defined in the baseline of DCASE 2021 Task 2\cite{Kawaguchi_arXiv2021_01},
during inference, we can also obtain an anomaly score from the classification confidence to make a decision for ASD, which is calculated by
\begin{equation}
  \mathcal{A}_{\text{aux}}(x_{i}^{t})= \log \frac{1-p_{\theta_{E}}\left(x_{i}^{t}\right)}{p_{\theta_{E}}\left(x_{i}^{t}\right)}
  \label{eq3}
\end{equation}
where $x_{i}^{t}$ represents the $i$-th test sample, and $p_{\theta_{E}}(\cdot)$ is the same as in Eq.(\ref{eq2}).


\begin{table*}[!ht]
  \scriptsize
  \centering
  \caption{Performance of AUC/$p$AUC(\%) (p=0.1) comparison on seven different ASD tasks.}
  \label{tab:results}
  \resizebox{\linewidth}{!}{
  \begin{tabular}{ccccccccc}
    \toprule
      Method & \textbf{ToyCar} & \textbf{ToyTrain} & \textbf{Fan} & \textbf{Gearbox} & \textbf{Pump} & \textbf{Slider} & \textbf{Valve} & \textbf{Average} \\
    \cmidrule(r){1-8} \cmidrule(r){9-9}
      AE\cite{Kawaguchi_arXiv2021_01} & 63.19/52.42 & 63.00/54.90 & 64.03/53.58 & 66.76/52.80 & 63.66/54.74 & 69.16/56.40 & 53.74/50.61 & 63.36/53.63 \\
      MobileNetV2\cite{Kawaguchi_arXiv2021_01} & 59.58/57.64 & 59.16/51.74 & 64.66/64.84 & 68.24/60.03 & 64.20/58.06 & 62.62/56.86 & 57.07/52.83 & 62.21/57.42 \\
    \cmidrule(r){1-8} \cmidrule(r){9-9}
      Tozicka\cite{TozickaNSW2021}	& \textbf{74.10}/\textbf{58.20}	& 62.00/49.60 &	57.80/51.90	& 73.90/58.90 & 56.90/53.50	& 64.70/54.30	& 55.10/52.10	& 63.50/54.07 \\
      Tan\cite{TanNTU2021}	& 60.76/53.20	& 60.56/54.35	& 64.58/52.84	& 68.37/52.75	& 64.88/55.23	& 68.45/56.04	& 54.64/50.82	& 63.18/53.60 \\
      Wang\cite{WangUCAS2021}	& 65.94/53.43	& 67.26/55.19	& 62.60/53.42	& 66.61/52.83	& 62.12/54.76	& 66.78/56.18	& 54.46/50.61	& 63.68/53.77 \\
      Yamashita\cite{YamashitaGifuUniv2021}	& 56.26/51.21	& \textbf{71.06}/57.14	& 60.04/52.77	& 60.57/52.17	& 54.50/52.65	& 67.55/58.11	& 60.00/54.38	& 61.43/54.06 \\

    \cmidrule(r){1-8} \cmidrule(r){9-9}
      AEGM-GAE & 69.73/56.84 & 67.83/55.17 & 75.19/59.21 & \textbf{74.28}/\textbf{63.52} & 65.07/60.75 & 72.65/61.35 & 57.86/52.21 & 69.03/58.44 \\
      AEGM-AC & 55.56/54.07 & 64.02/58.40 & 73.85/\textbf{70.53} & 59.44/54.32 & 68.63/60.93 & 69.17/61.75 & \textbf{77.90}/\textbf{63.65} & 66.94/\textbf{60.52} \\
      AEGM-ensemble & 64.76/57.14 & 68.41/\textbf{58.67} & \textbf{77.56}/66.19 & 71.14/60.02 & \textbf{70.08}/\textbf{61.49} & \textbf{75.00}/\textbf{63.12} & 74.76/56.25 & \textbf{71.67}/60.41\\
    \bottomrule
    \vspace{0.5em}
  \end{tabular}
  }
\end{table*}

\subsection{Group-based Decoders}
In conventional AEs with symmetric structure,
input samples are compressed and then reconstructed,
implicitly assuming that the input samples originate from a
single distribution.
However, in industrial settings, the situation is different. A machine typically contains different components (sections in this study). Noise and environmental variations in different operating states cause domain shifts between normal training samples. These samples often come from multiple distributions.
Therefore, in such cases, traditional AEs simply assume all input data come from a single distribution, leading to information loss and performance degradation in ASD.

Inspired by the masked autoencoder \cite{he2022masked},
we propose an asymmetric reconstruction AE with
group-based decoders to improve the model robustness against
domain shifts.
In this structure, as shown in Fig.\ref{fig:overview}(a),
$M$ decoders share the same encoder.
During training, we group data from the same section together and assign each decoder to one group.
Unlike the conventional AE model that assumes all input samples belong to the same distribution, this structure assumes data from different groups follow different distributions.
Here, each machine section corresponds to one specific acoustic domain.
The $M$ decoders retain as much distribution information
as possible in the training data, thus improving the model domain-specialization ability
and robustness.
In addition, this structure helps
establish clearer boundaries between distributions, boosting the performance of the proposed AC.

Furthermore, during model training, we aim to focus more on the
decoders that are harder to optimize and give larger gradients to
the corresponding ones. Thus, besides the group-based decoder
structure, we also propose an adaptive loss function $\mathcal{L}_\text{rec}$
to combine the $M$ decoders as
\begin{equation}
  \mathcal{L}_\text{rec}=\sum_{j=1}^{M} w_j\mathcal{L}_{\text{rec}_j}
  \label{eq6}
\end{equation}
where $\mathcal{L}_{\text{rec}_j}$ and $w_j$ are the reconstruction
loss of $j$-th decoder and its weight, respectively. And the
$\mathcal{L}_{\text{rec}_j}$ is defined as
\begin{equation}
  \mathcal{L}_{\text{rec}_j}=\sum_{i=1}^{N}  \Vert x_i-\mathcal{D}_{{\theta_{D}}_j}\left(\mathcal{E}_{\theta_{E}}\left(x_i\right)\right)\Vert_2^{2}
  \label{eq4}
\end{equation}
where $x_{i}$ represents the $i$-th input sample, $\mathcal{E}(\cdot)$ and $\mathcal{D}(\cdot)$ are the encoder and the $j$-th decoder of AEGM with parameters $\theta_{E}$ and $\theta_{{D}_j}$, respectively. The weight $w_j$ of each decoder's loss $\mathcal{L}_{\text{rec}_j}$ is defined as its proportion to the total loss and
calculated as
\begin{equation}
  w_j=\frac{\mathcal{L}_{\text{rec}_j}}{\sum_{j=1}^{M} \mathcal{L}_{\text{rec}_j}}
  \label{eq5}
\end{equation}

Given the $\mathcal{L}_\text{rec}$ and $\mathcal{L}_{\text{aux}}$, the proposed
AEGM is then jointly multi-optimized using the total loss as
\begin{equation}
  \mathcal{L}_\text{total}=\frac{\mathcal{L}_\text{aux}}{\mathcal{L}_\text{rec}+\mathcal{L}_\text{aux}} \mathcal{L}_\text{rec}+\frac{\mathcal{L}_\text{rec}}{\mathcal{L}_\text{rec}+\mathcal{L}_\text{aux}}\mathcal{L}_\text{aux}
  \label{eq8}
\end{equation}
With shared use of the encoder, this multi-task optimization makes the
bottleneck layer's embedding own strong abilities to perform
well for both the reconstruction and classification tasks.
On one hand, the model compresses the input information into the
bottleneck embedding through the reconstruction task.
On the other hand, the model maximizes utilization of the machine
section or domain-specific information, resulting in clear
classification boundaries in the latent space.

During inference, the machine section (ID, group, category) of each test sample $x_{i}^{t}$ is known in advance. We can thus compute the anomaly score $\mathcal{A}_{\text{rec}}(x_{i}^{t})$ using the reconstruction error of the corresponding decoder. Specifically, if $x_{i}^{t}$ belongs to section $j$,
its $\mathcal{A}_{\text{rec}}(x_{i}^{t})$ is computed as,
\begin{equation}
  \mathcal{A}_{\text{rec}}(x_{i}^{t})=  \Vert x_{i}^{t}-\mathcal{D}_{{\theta_{D}}_j}\left(\mathcal{E}_{\theta_{E}}\left(x_{i}^{t}\right)\right)\Vert_2 ^{2}
  \label{eq7}
\end{equation}
The above reconstruction error indicates the degree of abnormality, with higher values
indicating greater anomaly. Moreover, in our experiments,
we also explore combining $\mathcal{A}_{\text{rec}}(x_{i}^{t})$ and
$\mathcal{A}_{\text{aux}}(x_{i}^{t})$ to assess score ensemble performance.




\vspace{0.1em}
\section{Experiments}

In this section, we compare our experimental results with two official baseline systems and other reported results in DCASE 2021 Task 2\cite{Kawaguchi_arXiv2021_01}, using t-SNE\cite{van2008visualizing} visualization for detailed analysis.

\subsection{Datasets}

Note that we can only train with normal machine sounds because of the lack of anomalous sounds. We use ToyADMOS2\cite{harada2021toyadmos2} and MIMII DUE\cite{Tanabe_arXiv2021_01} datasets from DCASE 2021 Task 2 to build our ASD system AEGM.
These datasets contain seven types of normal and abnormal machine sounds,
consisting of a total of seven sub-datasets.
The abnormal conditions include leakage, rotation unbalance, bent, damage,
overload, etc., hard for humans to distinguish.
Each 10-second, single-channel sound clip is downsampled to 16 kHz.
The machine sounds are merged with real factory noise,
which presents a challenge for the task.
We use the development set for testing and each machine type
data in the development set is divided into three sections,
which are defined as subsets of each data.
Approximately 1000 normal segments can be used for training in each section,
and 100 normal and abnormal segments are used for testing.

\subsection{Experimental Setup}

As discussed in \cite{fang2022out}, it is challenging to
find a general approach for all types of anomaly detection,
so we use different acoustic features for different machines.
128-dimensional log-mel-spectrogram (log-mel) features are used for 
ToyCar, ToyTrain, Fan, Gearbox and Valve, while for the 
Pump and Slider, we use the 513-dimensional STFT spectrogram as 
acoustic features. All features are calculated with a frame size of 64ms with $50\%$ hop size.
In addition, for the tasks of Fan and Gearbox, we utilize 
a simple partially feature shuffle of
low-frequency components of input feature frames as data augmentation to enhance
the model robustness.

The models are trained for 600 epochs with
a batch size of 32, using the Adam optimizer.
The learning rate for GAE branch is set to 0.001 and 0.00001 is used for the AC branch. The structure of the AEGM encoder is shown in
Fig.\ref{fig:overview}(b), and each decoder has a symmetrical structure
as the encoder. As in the DCASE 2021 Task 2,
the area under the receiver operating characteristic curve (AUC)
and partial-AUC ($p$AUC) \cite{Kawaguchi_arXiv2021_01}, are used as our system evaluation metrics.


\subsection{Overall Results}

Table \ref{tab:results} shows the details of AUC/$p$AUC performance on seven
different ASD tasks, and the AE and MobileNetV2 \cite{sandler2018mobilenetv2} are two official baseline
results provided by the organizers of DCASE 2021 Task 2 \cite{Kawaguchi_arXiv2021_01}. AEGM-GAE
and AEGM-AC means we only take the anomaly score of the group-based
decoder or the auxiliary classifier to make the ASD decision, respectively.
AEGM-ensemble means that we apply mean-variance normalization
to both $\mathcal{A}_{\text{aux}}$ and $\mathcal{A}_{\text{rec}}$,
and then compute the final anomaly score as their mean. The
‘Average’ means taking the mean value of the AUC/$p$AUC performance for the seven machines. The other results are from other AE-based methods reported in the related technical reports.

As shown in the 2nd block of Table \ref{tab:results}, for the DCASE 2021 Task2 challenge, 
many attempts have been proposed to improve AE with
complex models but only achieve marginal gains or even performance degradation over the official baseline. For example, authors in \cite{TozickaNSW2021} used pretrained OpenL3 embeddings to train AE, and in \cite{WangUCAS2021}, the interpolation AE was utilized. They only obtain slight performance improvements. J. Yamashita in \cite{YamashitaGifuUniv2021}  employed VAE and IDNN, and E. Tan in \cite{TanNTU2021} used an ensemble of three AEs with different hyperparameters, leading to a small performance drop. This is because even with more powerful feature extractors or modified AEs, the ‘shortcut’ problem of AE is not addressed. Moreover, generic pre-trained models may not fit the ASD tasks well.

Our proposed AEGM method does not use complex network architectures but only pure MLP structures. By addressing the ‘shortcut’ problem with the group-based decoder and AC, it achieves competitive results. In terms of AUC, our method relatively improves over AE by 13.11\% and over MobileNetV2 by 15.20\%. This is due to the assistance of the AC and the group-based decoder structure, which maximally exploits the distribution characteristics of the data. The AC enables the latent layer of AEGM to have in-distribution classification ability, while the GAE reinforces this ability.

Interestingly, by comparing the results of AE and MobileNetV2 baselines, we see that the
AE outperforms MobileNetV2 in ToyCar, ToyTrain and Slider ASD tasks. On the other four tasks, MobileNetV2, which has a larger
model size than AE, performs better. Therefore, as the consistent observations in 
\cite{you2022unified}, we also find that it is difficult to build a unified
model that can perform well on all ASD tasks, because the
acoustic characteristics of different machines may be totally different.

Moreover, when comparing the results of the three anomaly score computation methods
of our proposed AEGM, the AEGM-GAE performs much better than the other two
on most ASD tasks, however, it is also unclear which one is the best for all tasks. For example, on the ‘Valve’ task, the AEGM-AC performs much better than all other systems. Moreover,
the ensemble method is also not the best, even it achieves slightly
better average performance than AEGM-GAE. This indicates that designing a unified ASD method remains an open challenge.


\subsection{Visualization of AEGM-AC}

It is worth noting that most methods perform
very poorly on non-stationary signals, such as the machine ‘Valve’.
However, it is interesting to observe that the AEGM-AC shows great improvement over other methods, the
AUC and $p$AUC reach 77.90\% and 63.65\%. It indicates that
the proposed AEGM-AC is effective for detecting anomalies
of non-stationary signals.
\begin{figure}[h]
  \centering
  \includegraphics[width=\linewidth]{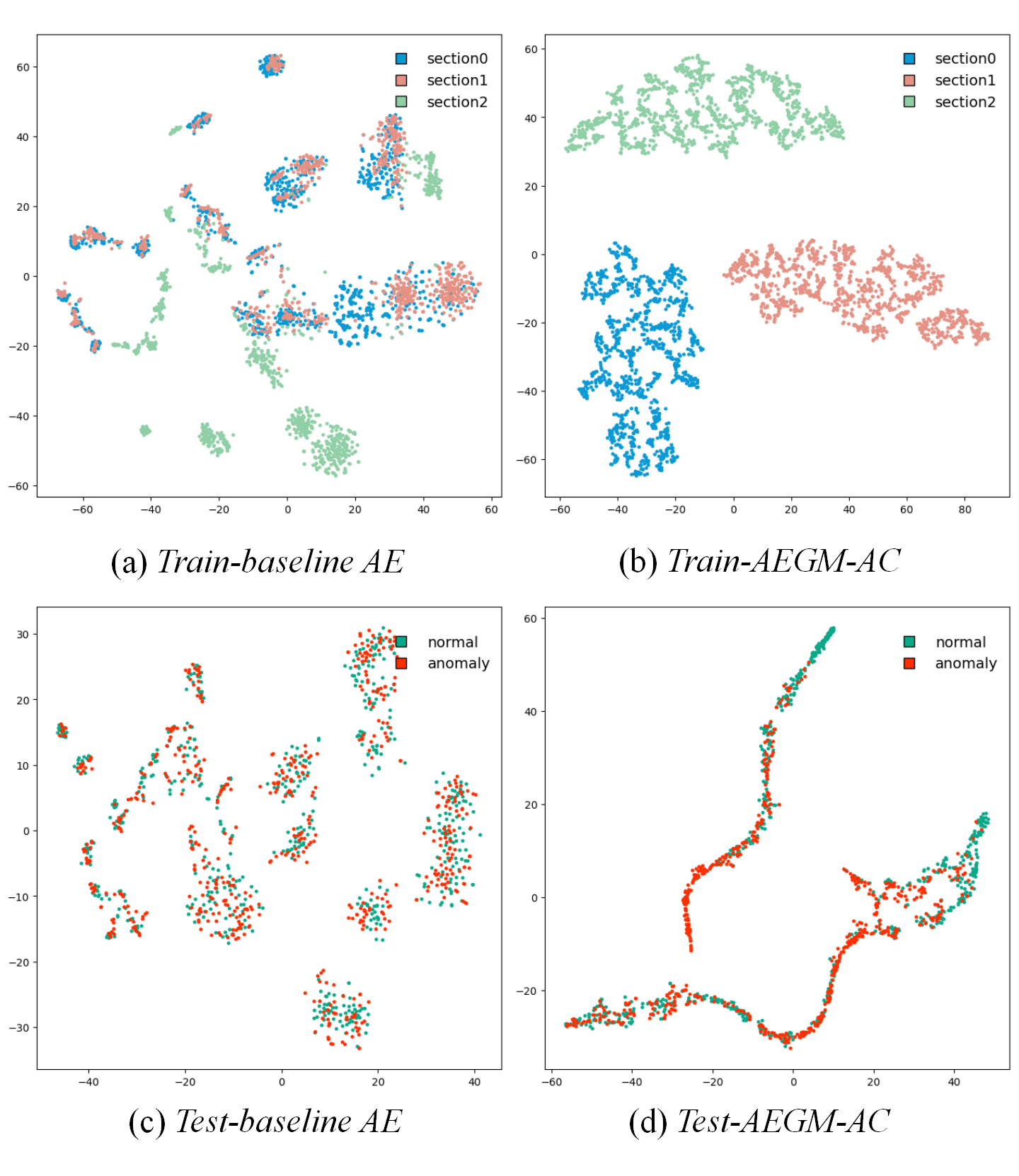}
  \caption{The t-SNE distribution of the training and test embeddings from the baseline AE and the proposed AEGM-AC for the machine ‘Valve’ ASD task.}
  \label{fig:tsne}
\end{figure}

Moreover, to make a more detailed analysis, in Fig.\ref{fig:tsne},
we plot a t-SNE \cite{van2008visualizing} to visualize the
discrimination ability between the official AE's bottleneck embeddings
and the latent embeddings of AEGM-AC on all the training
and test samples from machine ‘Valve’. Fig.\ref{fig:tsne}(a)(c)
are the visualization of official AE encoder's bottleneck
embeddings, while Fig.\ref{fig:tsne}(b)(d) are the t-SNEs on
outputs of the proposed AEGM-AC (the linear layer outputs of
the inserted AC).

Comparing Fig.\ref{fig:tsne}(b) with (a), it is clear that the overlap of
embeddings between different training sections are greatly reduced. It means that,
by just adding an AC branch with only one simple linear layer, together with the
group-based decoder structure, the discrimination
among the AC learned latent embeddings on the training samples is
strongly enhanced over the baseline AE model. Furthermore, when
we compare the distribution of Fig.\ref{fig:tsne}(d) with (c), the discrimination
of the proposed AEGM-AC is well generalized to the test set.
We see the embeddings of baseline AE are unable to
distinguish between normal and abnormal samples, while the AC generates
clear classification boundaries in the latent space,
effectively separating normal and abnormal samples.
This observation is very consistent with the one that is obtained
from Table \ref{tab:results} results on machine ‘Valve’.

\section{Conclusion}

In this study, we propose a new autoencoder framework AEGM to improve the
performance of industrial anomalous sound detection. Two improvements
are proposed to enhance the traditional reconstruction-based autoencoder.
A simple auxiliary classifier is first inserted to alleviate the `shortcut'
phenomenon of AE. Then, the group-based decoder structure with
adaptive loss function is designed to make the model learn better
domain-specific distribution in the normal training samples.
Experiments on the DCASE 2021 Task 2 show that our proposed method 
outperforms the official two baselines
significantly on almost all the seven anomalous sound detection tasks. In addition, we find that achieving consistent performance gains across seven diverse AD tasks remains difficult. 
Therefore, designing a unified ASD model will be the focus of our future work.

\bibliographystyle{IEEEbib}
\bibliography{isbib}

\begin{thebibliography}{10}

\bibitem{Kawaguchi_arXiv2021_01}
Yohei Kawaguchi, Keisuke Imoto, Yuma Koizumi, Noboru Harada, Daisuke Niizumi,
  Kota Dohi, Ryo Tanabe, Harsh Purohit, and Takashi Endo,
\newblock ``Description and discussion on {DCASE} 2021 challenge task 2:
  Unsupervised anomalous detection for machine condition monitoring under
  domain shifted conditions,''
\newblock in {\em Proc. {DCASE} 2021 Workshop}, 2021, pp. 186--190.

\bibitem{koizumi2018unsupervised}
Yuma Koizumi, Shoichiro Saito, Hisashi Uematsu, Yuta Kawachi, and Noboru
  Harada,
\newblock ``Unsupervised detection of anomalous sound based on deep learning
  and the neyman-pearson lemma,''
\newblock {\em IEEE/ACM TASLP}, vol. 27, no. 1, pp. 212--224, 2018.

\bibitem{huang2018internet}
Der-Chen Huang, Chun-Fu Lin, Chih-Yen Chen, and Jyh-Rou Sze,
\newblock ``The internet technology for defect detection system with deep
  learning method in smart factory,''
\newblock in {\em Proc. IEEE ICIM}, 2018, pp. 98--102.

\bibitem{marchi2015novel}
Erik Marchi, Fabio Vesperini, Florian Eyben, Stefano Squartini, and Bj{\"o}rn
  Schuller,
\newblock ``A novel approach for automatic acoustic novelty detection using a
  denoising autoencoder with bidirectional lstm neural networks,''
\newblock in {\em Proc. ICASSP}, 2015, pp. 1996--2000.

\bibitem{baldi2012autoencoders}
Pierre Baldi,
\newblock ``Autoencoders, unsupervised learning, and deep architectures,''
\newblock in {\em Proc. ICML Workshop Unsupervised Transfer Learn.}, 2012, pp.
  37--49.

\bibitem{kingma2013auto}
Diederik~P Kingma and Max Welling,
\newblock ``Auto-encoding variational bayes,''
\newblock {\em in Proc. ICLR}, 2014.

\bibitem{sabokrou2018adversarially}
Mohammad Sabokrou, Mohammad Khalooei, Mahmood Fathy, and Ehsan Adeli,
\newblock ``Adversarially learned one-class classifier for novelty detection,''
\newblock in {\em Proc. IEEE/CVF CVPR}, 2018, pp. 3379--3388.

\bibitem{hendrycks2018deep}
Dan Hendrycks, Mantas Mazeika, and Thomas Dietterich,
\newblock ``Deep anomaly detection with outlier exposure,''
\newblock {\em in Proc. ICLR}, 2018.

\bibitem{ruff2020rethinking}
Lukas Ruff, Robert~A Vandermeulen, Billy~Joe Franks, Klaus-Robert M{\"u}ller,
  and Marius Kloft,
\newblock ``Rethinking assumptions in deep anomaly detection,''
\newblock {\em arXiv:2006.00339}, 2020.

\bibitem{chen2020simple}
Ting Chen, Simon Kornblith, Mohammad Norouzi, and Geoffrey Hinton,
\newblock ``A simple framework for contrastive learning of visual
  representations,''
\newblock in {\em Proc. ICML}, 2020, pp. 1597--1607.

\bibitem{hojjati2022self}
Hadi Hojjati and Narges Armanfard,
\newblock ``Self-supervised acoustic anomaly detection via contrastive
  learning,''
\newblock in {\em Proc. ICASSP}, 2022, pp. 3253--3257.

\bibitem{purohit2020deep}
Harsh Purohit, Ryo Tanabe, Takashi Endo, Kaori Suefusa, Yuki Nikaido, and Yohei
  Kawaguchi,
\newblock ``Deep autoencodin {GMM}-based unsupervised anomaly detection in
  acoustic signals and its hyper-parameter optimization,''
\newblock {\em in Proc. {DCASE 2020 Workshop}}, 2020.

\bibitem{scott2004outlier}
David~W Scott,
\newblock ``Outlier detection and clustering by partial mixture modeling,''
\newblock in {\em Proc. Comput. Statist.}, 2004, pp. 453--464.

\bibitem{breunig2000lof}
Markus~M Breunig, Hans-Peter Kriegel, Raymond~T Ng, and J{\"o}rg Sander,
\newblock ``{LOF}: identifying density-based local outliers,''
\newblock in {\em Proc. ACM SIGMOD}, 2000, pp. 93--104.

\bibitem{liu2008isolation}
Fei~Tony Liu, Kai~Ming Ting, and Zhi-Hua Zhou,
\newblock ``Isolation forest,''
\newblock in {\em Proc. IEEE ICDM}, 2008, pp. 413--422.

\bibitem{an2015variational}
Jinwon An and Sungzoon Cho,
\newblock ``Variational autoencoder based anomaly detection using
  reconstruction probability,''
\newblock {\em Special Lecture on IE}, vol. 2, no. 1, pp. 1--18, 2015.

\bibitem{suefusa2020anomalous}
Kaori Suefusa, Tomoya Nishida, Harsh Purohit, Ryo Tanabe, Takashi Endo, and
  Yohei Kawaguchi,
\newblock ``Anomalous sound detection based on interpolation deep neural
  network,''
\newblock in {\em Proc. ICASSP}, 2020, pp. 271--275.

\bibitem{purohit2022hierarchical}
Harsh Purohit, Takashi Endo, Masaaki Yamamoto, and Yohei Kawaguchi,
\newblock ``Hierarchical conditional variational autoencoder based acoustic
  anomaly detection,''
\newblock in {\em Proc. EUSIPCO}, 2022, pp. 274--278.

\bibitem{georgescu2021background}
Mariana~Iuliana Georgescu, Radu~Tudor Ionescu, Fahad~Shahbaz Khan, Marius
  Popescu, and Mubarak Shah,
\newblock ``A background-agnostic framework with adversarial training for
  abnormal event detection in video,''
\newblock {\em IEEE TPAMI}, vol. 44, no. 9, pp. 4505--4523, 2021.

\bibitem{you2022unified}
Zhiyuan You, Lei Cui, Yujun Shen, Kai Yang, Xin Lu, Yu~Zheng, and Xinyi Le,
\newblock ``A unified model for multi-class anomaly detection,''
\newblock {\em in Proc. NIPS}, 2022.

\bibitem{pang2021deep}
Guansong Pang, Chunhua Shen, Longbing Cao, and Anton Van~Den Hengel,
\newblock ``Deep learning for anomaly detection: A review,''
\newblock {\em ACM Comput. Surv.}, vol. 54, no. 2, pp. 1--38, 2021.

\bibitem{TozickaNSW2021}
Jan Tozicka, Durkota Karel, and Linda Michal,
\newblock ``Unsupervised anomalous sound detection by siamese network and
  auto-encoder,''
\newblock Tech. {R}ep., DCASE2021 Challenge, July 2021.

\bibitem{TanNTU2021}
Ee-Leng Tan, Santi Peksi, and Duy~Hai Nguyen,
\newblock ``Anomaly sound detection using essemble of autoencoders,''
\newblock Tech. {R}ep., DCASE2021 Challenge, July 2021.

\bibitem{WangUCAS2021}
Shuo Wang, Zihao Li, Yuxuan Zhang, Kejian Guo, Shijin Chen, and Yan Pang,
\newblock ``Data augmentation and class-based ensembled cnn-conformer networks
  for sound event localization and detection,''
\newblock Tech. {R}ep., DCASE2021 Challenge, July 2021.

\bibitem{YamashitaGifuUniv2021}
Jun'ya Yamashita, Hayato Mori, Satoshi Tamura, and Satoru Hayamizu,
\newblock ``Vae-based anomaly detection with domain adaptation,''
\newblock Tech. {R}ep., DCASE2021 Challenge, July 2021.

\bibitem{he2022masked}
Kaiming He, Xinlei Chen, Saining Xie, Yanghao Li, Piotr Doll{\'a}r, and Ross
  Girshick,
\newblock ``Masked autoencoders are scalable vision learners,''
\newblock in {\em Proc. IEEE/CVF CVPR}, 2022, pp. 16000--16009.

\bibitem{van2008visualizing}
Laurens Van~der Maaten and Geoffrey Hinton,
\newblock ``Visualizing data using {t-SNE},''
\newblock {\em JMLR}, vol. 9, no. 11, 2008.

\bibitem{harada2021toyadmos2}
Noboru Harada, Daisuke Niizumi, Daiki Takeuchi, Yasunori Ohishi, Masahiro
  Yasuda, and Shoichiro Saito,
\newblock ``{ToyADMOS2}: Another dataset of miniature-machine operating sounds
  for anomalous sound detection under domain shift conditions,''
\newblock {\em in Proc. DCASE 2021 Workshop}, 2021.

\bibitem{Tanabe_arXiv2021_01}
Ryo Tanabe, Harsh Purohit, Kota Dohi, Takashi Endo, Yuki Nikaido, Toshiki
  Nakamura, and Yohei Kawaguchi,
\newblock ``{MIMII DUE}: Sound dataset for malfunctioning industrial machine
  investigation and inspection with domain shifts due to changes in operational
  and environmental conditions,''
\newblock {\em in Proc. IEEE WASPAA}, 2021.

\bibitem{fang2022out}
Zhen Fang, Yixuan Li, Jie Lu, Jiahua Dong, Bo~Han, and Feng Liu,
\newblock ``Is out-of-distribution detection learnable?,''
\newblock in {\em Proc. NIPS}, 2022.

\bibitem{sandler2018mobilenetv2}
Mark Sandler, Andrew Howard, Menglong Zhu, Andrey Zhmoginov, and Liang-Chieh
  Chen,
\newblock ``{MobileNetV2}: Inverted residuals and linear bottlenecks,''
\newblock in {\em Proc. IEEE/CVF CVPR}, 2018, pp. 4510--4520.

\end{thebibliography}


\end{document}